\documentclass[pre,twocolumn]{revtex4-1}
\usepackage{graphicx}
\usepackage{comment}
\usepackage{amsmath,amssymb,amsthm} 
\usepackage{bm}
\usepackage{verbatim}
\usepackage{float}
\begin{document}

		\title{ The role of the peptides in enzymes at the origin of live}

\author{S\o ren Toxvaerd }
\affiliation{DNRF centre  ``Glass and Time,'' IMFUFA, Department
 of Sciences, Roskilde University, Postbox 260, DK-4000 Roskilde, Denmark}
\date{\today}

\vspace*{0.7cm}

\begin{abstract}
	
   The peptides in biosystems are homochiral polymers of L-amino acids, but razemisate slowly
by an active isomerization kinetics.
The chemical reactions in biosystems are, however,  reversible and what  racemisates
 the peptides at the water activity in the biosystems
 can  ensure  homochirality at a smaller  activity.
 Here we show by a thermodynamics analysis and by  comprehensive Molecular Dynamics 
 simulations of models of peptides,  that the isomerization kinetics
 racemisates the
peptides at a high water activity  in
 agreement with experimental observations of aging of peptides
 , but
 enhances
 homochirality  at a smaller water activity.
 The hydrophobic
	core of the peptide  in an enzyme can ensure homochirality at a low water activity,
  and thus the establishment of homochirality at the origin of life and aging  of peptides and dead
 of biosystems might be strongly connected.

\end{abstract}

\maketitle

\section{Introduction}

 Biosystems consist of chiral polymers, where the building units are L- amino acids and D-carbohydrates.
 Another general  behavior of the polymers  in the cells, the peptides and DNA,
  is that they perform  "higher-order" conformational structures of the homochiral  units  \cite{Pauling,Watson,Levene}.  But
neither the peptides, nor DNA in the cells \cite{Lindahl,Modrich,Sancar} are long time stable. The
L-amino acids in aqueous solutions \cite{Bada1,Bada2},  and the  homochiral peptides
in the cells    racemisate slowly  \cite{Geiger,Fujii1,Hooi,Fujii2,Truscott1}. 
The racemisation of  the peptides in biosystems is associated with an increased concentration of water molecules in the peptides and
 lost $\beta$-sheet structure \cite{Fujii1}.
The water activities in biosystems differ with respect to the  extracellular- and  intracellular  water activity.
Whereas the  extracellular fluid in biosystems can be  characterized
as a rather ideal   solution ("saline solution"),  the intracellular fluid (cytosol) does not behave as a diluted  aqueous solution  \cite{Luby}.
The origin of homochirality  in the biosystems  is unknown, but since the homochirality  of the peptides  degenerates  in time,
is must have been established at a condition, which deviates from the physicochemical condition in biosystems.

\section{The  physicochemical condition for stable homochirality in aqueous solutions }

A pure racemic mixture of   D-  and L-molecules  can separate in homochiral domains  if the enthalpy gain,
due to the chiral discrimination, is bigger than the entropy of mixing, just as in the case of Pasteur's experiment \cite{Pasteur},
but now in a fluid state.
 A racemic mixture of chiral molecules and  without  an isomerization kinetics will  separate in homochiral subdomains  by molecular diffusion.
This more ordered state can spontaneously be obtained from a prebiodic state with racemic portions of the building units
 (sugars and amino acids)  
provided that the reaction Gibbs free energy, $\Delta_r G=\Delta_r H - T \Delta_r S<0$, is negative.

The separation to the homochiral states  has a negative reaction entropy, $\Delta_r S<0$.
If the  temperature, $T$, times the reaction entropy, $T\Delta_r S<0$  by a separation  from the racemic state  to the two homochiral states 
 is less negative than the corresponding reaction enthalpy,  $\Delta_r H<0$ it will  ensure a negative reaction Gibbs free energy.
A simple "ideal mixture" estimate of the  entropy contribution to the reaction Gibbs free energy  gives \cite{Tox1}
\begin{equation}
	T\Delta_r S \approx -  RTln2 \approx -  2 \;  kJ  mol^{-1}
\end{equation}
at moderate and biological relevant temperatures, and the system will phase separate into
homochiral domains,  if the gain of reaction enthalpy given by a "chiral discrimination" is strong enough, i.e.
\begin{equation}
 \Delta_r H< - 2   \;   kJ mol^{-1}.
\end{equation}
This fundamental physicochemical mechanism has recently be observed for
spontaneous phase separation in a  fluid mixture  into coexisting fluid domains of homochiral molecules  \cite{Dressel} . 

  An active isomerization kinetics
in a racemic mixture  with a reaction enthalpy,  $\Delta_r H< T\Delta_r S$  can, however, also  ensure 
the establishment of only one  homochiral domain \cite{Tox2}.   
Biochemical reactions are typically bimolecular.
The bimolecular  isomerization  kinetics of amino acids is   
  
\begin{equation}
\begin{array}{lllll}
 & E_{\mathrm{DD}}& & E_{\mathrm{DL}} \\
  \mathrm{D + D}& \rightleftharpoons
   &\mathrm{D + L}& \rightleftharpoons& \mathrm{L + L}\\
    & E_{\mathrm{DL}}& & E_{\mathrm{LL}} \\
    \end{array}
    \end{equation}
    between the two chiral species, D and L.
  The activation energy, $E_{\mathrm{DL}}$, for a  DL-collision
         which may  convert  a  D-molecule
          into  a L-molecule or \textit{vice verse},
          is, in a condensed racemic fluid,  less than the corresponding activation energy,  $E_{\mathrm{DD}}
             = E_{\mathrm{LL}}$, thus allowing a conversion  of one of the molecules in  the collisions.
The     inequality,
\begin{equation}
E_{\mathrm{DL}}<E_{\mathrm{LL}}= E_{\mathrm{DD}},
\end{equation}
accounts for the chiral discrimination with
 a lower potential energy   for a homochiral pair of molecules, than for an enantiomer pair, corresponding to a sufficient
 strength of the chiral discrimination.
 Let the rate constants, $ k_{\mathrm{DL}}$ and  $  k_{\mathrm{LL}}=k_{\mathrm{DD}}$, be given by
 Arrhenius expressions:
$k_{\mathrm{DL}}= A_{\mathrm{DL}}exp^{E_{\mathrm{DL}}/RT},
k_{\mathrm{LL}}= A_{\mathrm{LL}}exp^{E_{\mathrm{LL}}/RT}.$
 The difference in activation energies is  proportional to the (local) chiral discrimination, 
$E_{\mathrm{DL}} -E_{\mathrm{LL}}  \propto \Delta_r H(x)$,
where a local gain in enthalpy, $-\Delta_rH(x)>0 $, by a conversion of the stereo configuration  is a function of the
$\textit local$ composition, $x({\bf r})$ of the chiral  particles  nearby the position  ${\bf r}$.

The strength of chiral discrimination depends on how well the molecules fits into   a chiral structure and is given
by the  complex potential function between  a molecule
and its chiral  neighbours.
 Primarily   $\Delta_r H(x) $
depends on the excess number of homochiral neighbours by a change of a configuration.
 Consider a simple example:
let a molecule  (or a chiral unit in a peptide)  at the position ${\bf r}$  be in e.g. a L configuration before it is activated to  a  (intramolecular) transition state configuration.
 It will with a Boltzmann probability
choose  the configuration with lowest  potential energy. A  molecule in a  liquid mixture has typically
 eleven to twelve nearest neighbours. Let e.g. five of them be
in a D configuration, four of the neighbours in a L configuration and three neighbours be  indifferent
water molecules. It corresponds to a
local racemic  composition  near  ${\bf r}$    of thirteen molecules with an equal amount of D and L
 configurations before the activation of   the  L molecule at ${\bf r}$ . But 
due to the local excess of D molecules around the  activated molecule, it
  will most likely  turn into a D configuration by which the
system tends to a lower energy, but now with an excess of D configurations. Thus
a strong chiral discrimination  together with an  isomerization kinetics will  ensure a  separation
of a racemic mixture on a molecular level and
 tend to order the chiral units in homochiral clusters and subdomains. 
The same kinetics for homochirality must also be valid for  the chiral units 
 in  the peptides,  where the symmetry break is caused by the chirality in the domain of nearest chiral units.

The homochiral dominance   are obtained by, what could be expressed as
{\textit{self-stabilizing chance}} \cite{Tox1}. The deviation from a racemic mixture is self-stabilizing, because
  homochiral clusters- or sub domains catalyze their own growth by the isomerization kinetics,
 which mainly takes place in the interface, whereas  the chiral discrimination
 inside the homochiral domains slows down the kinetics and the conversions of the configurations, which always are unfavorable.
 Still one needs to explain the observed dominance, since the kinetics
seems only to enhance the separation, but $\textit{it does not  favour one of the chiral}$
$\textit{ species}$.
The break of symmetry  on a macroscopic scale, and the establishment of one stable homochiral domain will appear, when one of
the homochiral domains encapsulates the other domain \cite{Tox2,Tox3}.

The simple carbohydrates and amino acids are soluble in water.
 The chiral discrimination
in an $ideal$  solution of amino acids or simple carbohydrates  in water with the mole fraction  $x_{w}$, is reduced proportional
to the number of chiral neighbours, and is approximately 
\begin{equation}
	\Delta_r H(aq) \approx (1-x_{w}) \Delta_r H.
\end{equation}
 The chiral discrimination  near a surface,  $\Delta_r H_{surf.}(aq) \approx 0.5(1-x_{w}) \Delta_r H, $ is further reduced by a
factor of  the order two due to  halving of the number of nearest neighbours,
 and with the catastrophic result, that  a  chiral purification  is only possible
for an extreme high concentration of the chiral units or an extreme high strength of chiral discrimination, no matter where in the fluid.
It explains why the presence of water molecules   affects the quatenary structure of a peptide and reduces the chiral order \cite{Fujii1,Fujii2}.

 The stability of chiral order in presence  of water can be obtained for ideal mixtures.
The equilibrium constant $K$  for  a diluted solution 
  of e.g. L-chiral molecules or a  peptide with    L- amino units in equilibrium with a small fraction of its enantiomer
  by an active isomerization kinetics, is 
\begin{eqnarray}
 K= \frac{x_{\mathrm{L}}x_{\mathrm{D}}}{x_{\mathrm{L}}^2}   = \frac{x_{\mathrm{L}}(1-x_{\mathrm{L}}-x_w)}{x_{\mathrm{L}}^2}  \nonumber \\
 = \frac{k_{\mathrm{LL}}}{k_{\mathrm{DL}}}= exp^{(E_{\mathrm{LL}}-E_{\mathrm{DL}})/RT}=exp^{x_{w} \Delta_r H /RT}.
\end{eqnarray}
The  equation determines the mole fractions  $x_{\mathrm{L}}$ and $x_{\mathrm{D}}$ as a function of $x_{w}$ and $\Delta_r H$
 for an aqueous solution of chiral molecules in equilibrium with a small fraction of
its image molecules by an isomerization kinetics.

\section{ Molecular Dynamics simulations of peptides in aqueous solutions and with isomerization kinetics }

  The chiral discrimination, $\Delta_r H$,
  depends on  how well a chiral molecule, or a mirror image of the molecule, packs with other chiral
  molecules (e.g.  amino acids or  carbohydrates).
The net energy difference is given
by   complex potential functions.
But since it is the  net gain of energy  which gives the strength,  it can be obtained
 by an  energy function, which ensures a correct  gain of energy  from the
  interactions between the  molecule and its neighbour molecules. 
Here we simulate, by Molecular Dynamics (MD), such systems of peptides  of "united atom"  units  with chiral energy differentiation.

The system consists of $N=40000$ Lennard Jones (LJ)
particles in a cubic box with periodical boundaries \cite{ToxMD}. The MD simulations
are performed with the central difference algorithm in the leap-frog
 version, and the forces for particle distances greater than $r_{cut}$ are ignored. There are different ways
to take the non-analyticity of the force at $r_{cut}$  into account. The most stable and energy conserving way
is to cut and shift the forces (SF), \cite{Tox4} by which one avoids a  nonphysical force gradient at the cut.
 The SF-MD simulations are performed for a temperature $T$=1.00 and a density $\rho=$0.80, which corresponds to
a condensed liquid at a moderate ("room")  temperature.

  The  peptide chains  are constructed by linking LJ-units together
 by  reflecting the  LJ potential between two neighbour units in the chain
  at the potential minimum \cite{Akkermans}.
This  anharmonic bond potential is LJ-like and ensures a smooth interaction of an "amino acid unit"
in the peptide with the water particles,  as well as with the other  units in the chain molecule.

\subsection{ Potentials for hydrophobic and hydrophilic behaviour}

The structure of a fluid is  mainly determined by the forces within the first coordination shell  of
nearest neighbour particles  \cite{Tox5}, and it is also
the short range  attractive forces which determines the strength of the chiral discrimination.
 At the state point  $(T, \rho)=$(1.00,0.80) 
 the range of the first coordination shell (fcs) in the LJ system is $r_{\mathrm{frc}} \approx 1.55$.

  The range of attraction  for homochiral pairs (DD) or (LL)   is taken to be  equal to
 the radius of the first coordination shell, $r_{cut}(\mathrm{DD})= r_{cut}(\mathrm{LL})=r_{\mathrm{frc}} = 1.55$,
 by which one obtains a maximum attraction between homochiral pairs.
 A smaller  mean energy for a racemic composition is then achieved by  using a smaller range of attraction between enantiomers.
The range of attraction between two enantiomers is  taken to be  $r_{cut}(\mathrm{DL})$=1.35, by which the mean potential energy
difference, $\Delta u$,  between a racemic mixture and a homochiral fluid is determined to be  $\approx T ln 2$.
 (The MD is for canonical ensemble dynamics (NVT), where the chiral
discrimination is given, not by an  enthalpy difference , but with the corresponding  potential energy difference.)
Consistent with this choice,  MD simulations of a racemic mixture without isomerization kinetics  separate slowly into a D- and a L- reach domains for a smaller
range, $r_{cut}(\mathrm{DL}) < 1 .35$, as one shall expect from  thermodynamics considerations \cite{Tox4}.

Molecules can be sorted into hydrophilic- and hydrophobic molecules according to their solubility in water.
But the word "hydrophilicity" is perhaps a bit misleading, since there is only one molecule which is hydrophilic, and
this is $\mathrm{H}_2\mathrm{O}$. And although all simple carbohydrates and amino acids are soluble in water  for small or moderate
concentrations, they separate at higher concentrations.
   The strong hydrophilicity  between LJ "water" molecules (W) is achieved by using  $r_{cut}(\mathrm{WW})=1.55$, i.e. 
a strong
mean attraction between pairs of water molecules, equal to the strong  attraction between two homochiral units. 

 The hydrophilicity or hydrophobicity between  a water molecule  and a chiral molecule
-or chiral unit in a peptide,  is also monitored by the range of attraction.  A hydrophobic D or L unit
have a  small attraction (small water activity) to a water molecule. This hydrophobicity
 is achieved by using a cut  $r_{cut}(\mathrm{DW})=r_{cut}(\mathrm{LW})
 \leq 1.35$, i.e. less or equal to the attraction between two enantiomers, whereas 
 a more hydrophilic unit in the peptide have an attraction  $r_{cut}(\mathrm{DW})=r_{cut}(\mathrm{LW}) \geq 1.40$,
which corresponds to a higher water activity.

We have constructed chains with different numbers, $N_p$,  of  LJ-units  and  in aqueous solutions with $N_w=N-N_p$ water molecules.
The hydrophobic peptides  have a compact globular form with a low fraction of water molecules
in the peptides, whereas the chains swell up at a higher water activity.

\subsection{ Aqueous solutions of chiral molecules with  isomerization kinetics  }

 The isomerization kinetics is performed as described in Ref. 6.
A particle, No. $i$, at time $t$ is activated by a collision with one of its nearest neighbours, No. $j$, if the potential energy
at the collision
\begin{equation}
u_{ij}(t) \geq E,
\end{equation}
where $ E$ is equal to  $E_{\mathrm{DD}}=E_{\mathrm{LL}}$,  if $j$ has the same chirality as $i$,  and equal to
 $E_{\mathrm{DL}}$ if not.

 The (total) potential energy
 of particle No $i$,   at the time where it collides with $j$, is
\begin{equation}
	u_i(t) =\frac{1}{2} \Sigma_{k} u_{ik}(t)
\end{equation}
for the  sum over interactions with $i$'s  nearest neighbours. Correspondingly the potential energy of $i$ is
\begin{equation}
	\tilde u_i(t) =\frac{1}{2} \Sigma_{k} \tilde u_{ik}(t),
\end{equation}
if $i$'s chirality is changed.
    The chirality of particle No. $i$  is  then changed   in the
traditional way by a Boltzmann probability from $\Delta_r u=\tilde u_i(t) - u_i(t) $ \cite{Tox2}. 

According to the thermodynamics, the strength of the chiral discrimination  can be obtained from the activation energies,
$E_{\mathrm{DL}}$ and $E_{\mathrm{LL}}=E_{\mathrm{DD}}$.  The chiral discrimination in favor of homochirality has to be 
\begin{equation}
 -\Delta_r H= E_{\mathrm{LL}}-E_{\mathrm{DL}} \geq T \Delta_r S=T ln 2.
\end{equation}

The MD systems perform symmetry breaks with chiral purification for
 $E_{\mathrm{LL}}-E_{\mathrm{DL}}=E_{\mathrm{DD}}-E_{\mathrm{DL}} \geq 2T$,
 in agreement with the thermodynamics (ideal mixture) estimate \cite{Tox1}.
The isomerization kinetics in the next Section are for $E_{\mathrm{LL}}-E_{\mathrm{DL}}=E_{\mathrm{DD}}-E_{\mathrm{DL}}= 8T-5T$.

\section{ Peptides in aqueous solutions}
 
 The simulations of amino acids and peptides in aqueous solutions confirm  the thermodynamic derivations in
 Section II.  It is only possible to obtain a symmetry break  and chiral dominance in the MD system for concentrated 
solutions of chiral molecules, here  with $x_w \leq 0.20$ (Figure 1). But although all simple carbohydrates and amino acids are soluble in water  for small or moderate
concentrations, they separate  and crystallize  at higher concentrations.  Biosystems are, however in a fluid state, and there seems to be
only one possibility for a selforganised chiral purification and maintenance of homochirality in aqueous solutions:
$\textit{peptides with a hydrophobic core. } $

\begin{figure}
\begin{center}
\includegraphics[width=6cm,angle=-90]{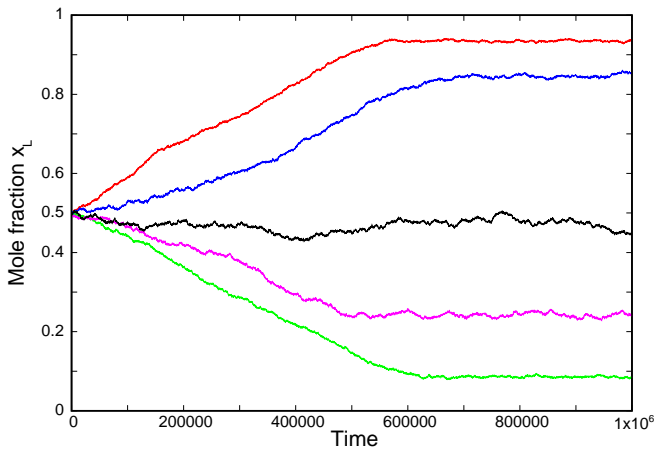}
\end{center}
\caption{  Mole fraction $x_{\mathrm{L}}(t)$  as a function of time (steps)  in aqueous  solutions of simple chiral
	molecules (e.g. amino acids)  with isomerization kinetics and a strong chiral discrimination.
	The start configurations are racemic ($x_{\mathrm{L}}(0)=0.5$). Red curve is
        $x_{\mathrm{L}}(t)$ for a solution with     a   mole  fraction of water molecules
 $x_{w}=0.025$; green curve: $x_{w}=0.05$; blue curve:  $x _{w}=0.125$;
  magenta curve: $x_{w}=0.2$ and  black curve: $x_{w}=0.25$.
The isomerization kinetics does not favor one of the chiral conformations,
 and a L or  D  dominance is obtained by chance with equal
probability, as demonstrated by the examples  in the figure. The equilibrium
 mean fractions after the symmetry breaks are given by Eq. (6).}
\end{figure}

\begin{figure}
\begin{center}
\includegraphics [width=6cm,angle=-90]{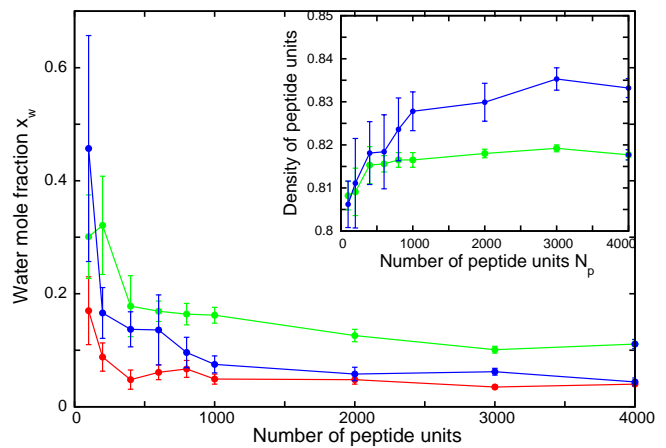}
\end{center}
\caption{ Mole fraction, $x_w$,  of water molecules in the sphere with radius equal to the radius of gyration for  peptides
with  $N_p$  hydrophobic units with strong chiral discriminations.
	The red points are for  homochiral peptides  and the green points are for racemic peptides.
The blue points are for  hydrophobic peptides  with isomerization kinetics and they have only  chiral dominance for $N_p \geq 400$. The inset
 shows the corresponding densities of the units and for  racemic peptides (green) and peptides with isomerization kinetics (blue).
 (The uncertainties are obtained from  ten independent simulations.)}
\end{figure}

We have constructed chains with different numbers, $N_p$,  of  chiral hydrophobic or hydrophilic  units,
  and  in aqueous solutions.
The hydrophobic peptides  have a compact globular form with a low fraction of water molecules
in the peptides, whereas the chains swell up at a higher water activities. 
There are also differences in the conformations for chains with a  racemic (random D- and L-) and a homochiral composition of hydrophobic units.
The racemic peptides (Figure 2, green points) have a significant higher water content
than the homochiral peptides (red points) , and it exceeds $x_{w} \approx 0.20$ for
the  racemic peptides with chain lengths less than $N_p \approx 400$. The blue points are for  hydrophobic peptides
with isomerization kinetics. These properties shown in Figure 2  are   in qualitative agreement with the  experimental
observation of a peptide  after aging with loss of homochirality \cite{Fujii1}.

\begin{figure}
\begin{center}
\includegraphics[width=6cm,angle=-90]{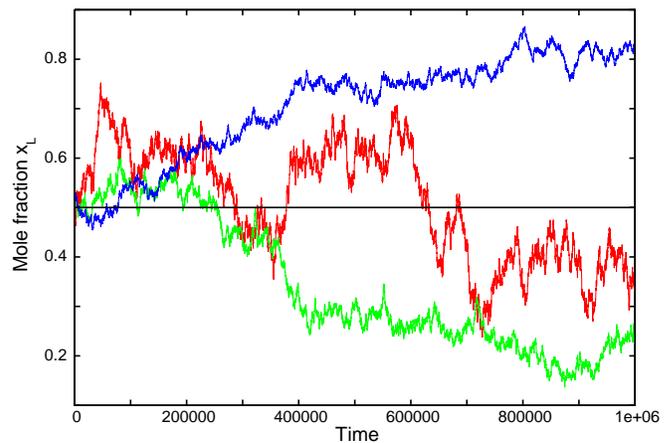}
\end{center}
	\caption{ Mole fraction $x_{\mathrm{L}}$ of L-units in  chains with $N_p=400$ (red), $N_p=1000$ (green) and  $N_p=2000$ (blue) chiral units, and with
isomerization kinetics.
 The chains have a  random and racemic distribution
at the start.}
\end{figure}

\begin{figure}
\begin{center}
 \includegraphics[width=6cm,angle=0]{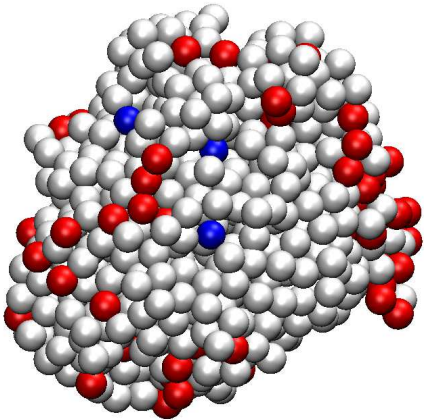}
\end{center}
\caption{ A peptide  of $N_p=1000$   hydrophobic units and
with  isomerization kinetics. The peptide is compact and rather homochiral ($x_{\mathrm{L}}$=0.15)  with dominating  D-units (white). The
L-units (red) are mainly located in the water-peptide interface. There are  15  water molecules (blue), mainly located
 in pockets of the compact peptide and with a few  in the interior. }
\end{figure}

\begin{figure}
\begin{center}
\includegraphics[width=5cm,angle=0]{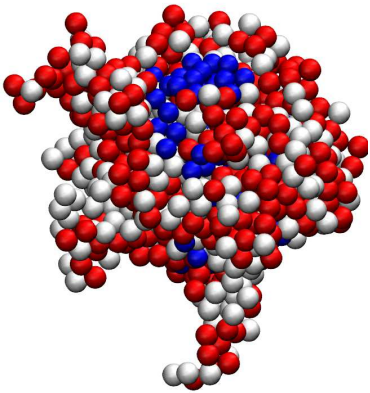}
\end{center}
\caption{ The peptide with  $N_p=1000$ units at a higher water activity,  and with  isomerization kinetics. The composition  (L/D=red/white) is now  $\approx$ racemic and the concentration of the water molecules are increased to 39.}
\end{figure}

Figure 3 shows the evolution of chiral dominance for different sizes of the hydrophobic  peptides.
The presence of water with a mole fraction $x_w \geq 0.20$ makes the homochiral state unstable (Figure 1), 
and favor the racemic state according to the thermodynamics, and the results in Figure 2
(blue line and points) shows that
small peptides with hydrophobic units and isomerization kinetics  contain water molecules with a fraction, which exceeds this stability limit.
In accordance with these results we observe, that  peptides of hydrophobic units  with isomerization kinetics and chiral discriminations
  can not maintain  homochirality when   $N_p \leq 400$ , 
and  Figure 3 demonstrates this fact. The  fraction of the L-state, $x_{\mathrm{L}}$, for a   small peptide of  $N_p = 400$ units (red line)
  fluctuates, but
remains racemic in mean. For longer chains one obtains, however, a symmetry break to a state
with a dominating chirality, either to a D-state ( $N_p=$1000: green), or  to a  L
-state ($N_p=$2000: blue).

The   hydrophobic peptide with $N_p=1000$ units, with $x_{\mathrm{L}}$(t) shown in Figure 3 (green line) is rather compact and globular.
The peptide was simulated over a longer time period, where it maintains  the chiral D dominance. The configuration at the end of the
simulation is shown in Figure 4. The strong hydrophobicity corresponds to a small water activity. If this peptide, however is exposed
to a higher water activity it swells up with a bigger contents of water in the core, and  the peptide looses its homochirality (Figure 5).

We have performed  many simulations,  for different strengths of hydrophobicity, water activity and
chiral discrimination ($E_{\mathrm{DL}}-E_{\mathrm{LL}}$) , and they conform the result, shown in the figures, that spontaneous chirality is only
achieved for sufficient strength of chiral discrimination, $E_{\mathrm{DL}}-E_{\mathrm{LL}}=E_{\mathrm{DL}}-E_{\mathrm{DD}} \geq 2$, and for hydrophobic peptides
with a molfraction of water less than $x_w \approx $ 0.20, 
 in accordance with the thermodynamics for obtaining spontaneous homochirality.
\section{Perspective}
 The strength of chiral discrimination
  depends on  how well a molecule packs with   a chiral  molecule  or  with  the mirror form of the  molecule.
Normally one associates it with how well a molecule fits with a copy of itself,
 compared with how well it fits with  its mirror image molecule.  But there is also chiral discriminations between different amino acids
 \cite{Viedma, Tarasevych}
 and between simple carbohydrates and amino acids  \cite{Kock,Takats,Breslow}.
There is  a chiral discrimination between  a simple carbon hydrate, D-Glyceraldehyde, and the amino acid  L-Serine \cite{Kock,Takats},
which can explain the overall dominance of D-carbohydrates and L-amino acids in biosystems.

The present investigation indicates that the  function  of the peptides in enzymes,  at the origin of homochirality and life, first were to stabilize
homochirality 
of the  units in the hydrophobic core of the peptides, whereby they
act as the "backbones" with a stereo specific  surfaces for the homochiral purification of carbohydrates and amino acids
by the isomerization kinetics.
The chemical reactions in biosystems are reversible, and what ensures
 homochirality at low water activity will racemisate at higher water activity.
The water activity at the establishment of homochirality must necessary have been  somewhat smaller than the 
aqueous cytosol solutions in the cells. The smaller water activity can be obtained by a more concentrated solutions
of amino acids and carbohydrates,
 and at  somewhat higher  ionic concentrations than the  ionic concentrations in the cytosol. 
Experiments on such aqueous solutions  can reveal, whether it is possible to obtain spontaneous chiral purification and to maintain homochirality in
 presence of peptides with  hydrophobic cores.

				\section{Acknowledgment}
The centre  $\textit{Glass and Time}$, Department of Science and Environment, Recoiled University is
gratefully acknowledged.

{}

\begin{thebibliography}{99}
\bibitem{ToxMD} In MD the mass $m$ is usually included in the time unit.
 The unit length, energy and time used for LJ systems are, respectively,
 $\sigma$, $\epsilon$ and $\sigma\sqrt{m/\epsilon}$.
\end{thebibliography}
\end{document}